# Gridscape II: A Customisable and Pluggable Grid Monitoring Portal and its Integration with Google Maps

Hussein Gibbins and Rajkumar Buyya

**Gri**d Computing and **D**istributed **S**ystems (GRIDS) Laboratory
Department of Computer Science and Software Engineering
The University of Melbourne, Australia
{hag, raj}@csse.unimelb.edu.au

## ABSTRACT

**Grid computing has emerged as an effective means of facilitating the sharing of distributed heterogeneous resources, enabling collaboration in large scale environments. However, the nature of Grid systems, coupled with the overabundance and fragmentation of information, makes it difficult to monitor resources, services, and computations in order to plan and make decisions. In this paper we present Gridscape II, a customisable portal component that can be used on its own or plugged in to compliment existing Grid portals. Gridscape II manages the gathering of information from arbitrary, heterogeneous and distributed sources and presents them together seamlessly within a single interface. It also leverages the Google Maps API in order to provide a highly interactive user interface. Gridscape II is simple and easy to use, providing a solution to those users who don't wish to invest heavily in developing their own monitoring portal from scratch, and also for those users who want something that is easy to customise and extend for their specific needs.**

## 1 INTRODUCTION

A pervasive problem seen in personal information management is information fragmentation [1]. This problem is also present in Network and Grid system information management, where information may be fragmented by physical location, device or even by the various tools designed to help manage it. Coupled with the current overabundance of information, this problem severely hinders people's ability to make intelligent decisions and take appropriate actions, due to the inability to easily locate and interpret information.

Grid computing [2] has emerged as a new paradigm for sharing distributed heterogeneous resources to facilitate collaboration in environments such as in enterprises [3] and e-Science [4] applications. The resources that make up these Grids are diverse and include scientific instruments, computational resources, application services, and data stores. Information services, such as Globus' Monitoring and Discovery System (MDS) [5] for compute resources or the Storage Resource Broker's Meta Information Catalog (SRB MCAT) [6], from the San Diego Supercomputer Center (SDSC), for data stores, are a key part of Grid middleware, providing fundamental mechanisms for monitoring, and hence for planning and adapting application behaviour. However, being able to monitor resources, services, and computations is challenging due to the heterogeneous nature, large number, dynamic behaviour, and geographical distribution of these resources. Each system may adhere to different standards and provide information based on various schemas and policies, and it therefore becomes hard to manage. However, users need a tool that presents concise information about available resources possibly in a single interface without the need of dealing with and swapping from a range of tools when checking the status of their resources. Also, every organisation need not invest heavily in developing or deploying Grid monitoring solutions for their own Grids.

At the very minimum, resource information can be gathered by directly querying resources or middleware specific information services. However, this may often require manually querying a number of different systems in order to discover pieces of information. This presents the need for information management tools to assist in gathering and aggregating information and simplifying the task of extracting required information. The ways in which individuals and groups organise and work with data also differ [7], meaning flexible approaches that can be tailored to individual needs are required.

Gridscape II, presented in this paper, aims to minimize this problem by assisting in the creation of Web based portals as well as easing tasks such as administering these portals, and thus helping users to find the information that is important to them. Web portals have become a popular way of providing a single interface to multiple different systems or fragments of information. The main design premises of a Grid monitoring portal such as Gridscape II are that it should:

- Manage diverse forms of resource information from various types of information sources;
- Allow new information sources to be easily introduced;
- Allow for simple portal management and administration;
- Provide a clear and intuitive presentation of resource information in an interactive and dynamic portal;
- Have a flexible design and implementation such that core components can be reused in building new components, presentation of information can be easily changed and a high level of portability and accessibility (from the web browser perspective) can be provided.

The remainder of this paper is structured as follows. Section 2 presents a discussion on the strengths and weaknesses of some currently existing solutions, identifying some unsatisfied requirements and areas of improvement to be addressed by Gridscape II. Sections 3 and 4 discuss how Gridscape II fits into the Grid architecture and provide details of its design and implementation. Section 5 illustrates how Gridscape II is used in practice. Finally, Section 6 presents future work and concludes the paper.

## 2   RELATED WORK

Currently, there are a number of solutions that aim to assist in the monitoring of Grid systems. Some of these are successful in their ability to provide detailed low-level system information while the strengths of others are their flexibility and adaptability. Here we discuss a few representative implementations and see how each addresses or fails to address certain issues.

The Grid Monitor of the Advanced Resource Connector (ARC), formally known as the NorduGrid Grid Monitor [8], addresses the problem of providing low level information by processing information provided by MDS and other services and presenting that in a user-friendly interface. One downside to this implementation however, is that it is tailored specifically to the ARC project and is not available to be used to monitor Grid resources other than those that are part of the ARC project publishing information to ARC Information System. Ganglia [9], is a distributed monitoring system that provides detailed information about resources. Unlike the ARC Grid Monitor, Ganglia is a flexible framework that can be utilised to monitor many types of high-performance computing systems including clusters and Grids. However, there is one caveat, that in order to achieve high levels of scalability and to provide such detailed resource information, Ganglia relies on Ganglia-specific daemons that run on cluster nodes, collect information and send multicast packets to other entities in the system. The problem with this approach is that considerable effort and administrative overhead is required to set up such systems and there are also security issues and policies with resources across various administrative domains where resource owners may resist installation or have preference for other monitoring solutions. MonALISA [10] follows a similar approach to Ganglia and provides a comprehensive monitoring solution with a distributed architecture; however it also suffers from the problem of needing to install monitoring agents on each resource. One of the strengths that MonALISA possesses that the above mentioned systems ignore is an extremely informative user interface that includes a geographical view of resources, making it easy to observe the statuses of large collections of resources at a glance. This improves the user experience and the ability to quickly identify and navigate for required information. A problem with the MonALISA client is that its comprehensive client is a stand-alone application that needs to be downloaded and installed on the client computer, reducing its accessibility. GridCat [11] is a Web portal based monitoring system that includes a map to visualise resources and includes detailed Grid information including job status, however it doesn't offer a simple means for extending its functionality beyond its current capability, which includes support for Globus, LCG and TerraGrid resources.

More flexible and adaptable solutions are toolkits that assist in the development of Grid portals and work with various types of middleware and information sources. Examples of such toolkits are GridPort [12] and GridSphere Grid Portlets [13]. These toolkits use portlet technology for reusability and aim to support all aspects of Grid systems including job execution, management and resource monitoring. GridPort, which at the time of writing is no longer being actively developed, focuses on providing a toolkit to simplify access to heterogeneous Grid services through a single API. Using their Grid Portal Information Repository (GPIR) architecture, all information is restricted to a common set of attributes of specific schemas in a number of categories such as: jobs, load, status and network latency. This limitation leads to a loss of richness of information through generalisation. Grid Portlets, an extension of the GridSphere portlet framework, provides a service for monitoring Grid resources and supports customisation for multiple different information sources to coexist. The limitations of this implementation are that there is no single collective view of all resources, no geographical visualisation (map interface), and the process of gathering resource information can only be run within a web server thus restricting its flexibility. Also, while these two approaches use portlet technology, they are small components dependant on larger systems. MapCenter [14] is another tool to help create portals with a similar approach and design methodology to Gridscape II. MapCenter however is not portlet based and instead generates static HTML at certain poll intervals. Also, Gridscape II has the ability to query any resource type and customise the presentation of that information within a single view, which is unavailable in MapCenter.

**Gridscape II Contributions**: Both the previous implementation of Gridscape [15] and its successor, the subject of this paper, aim to provide a high-level, user-friendly and highly customisable portal interface in order to present the status of Grid resources. Both also interact with existing technology so that no additional installation or configuration of Grid resource is required. Major improvements over the previous implementation of Gridscape II are that it supports the integration of multiple arbitrary information sources through an extensible design; it provides a simple customisation mechanism to allow it to be enhanced to meet the specific needs of each individual Grid portal. Other improvements are integration with Google Maps, simplified portal administration and the use of portlet-based web components which means it can be plugged into other Grid portals to compliment them.

## 3   ARCHITECTURE

The Gridscape II architecture consists of four main components:

1. Gridscape II Core
2. Gridscape II Resource Monitor
3. Gridscape II Portal
4. Interactive Client-Side Map (Google Maps Interface)

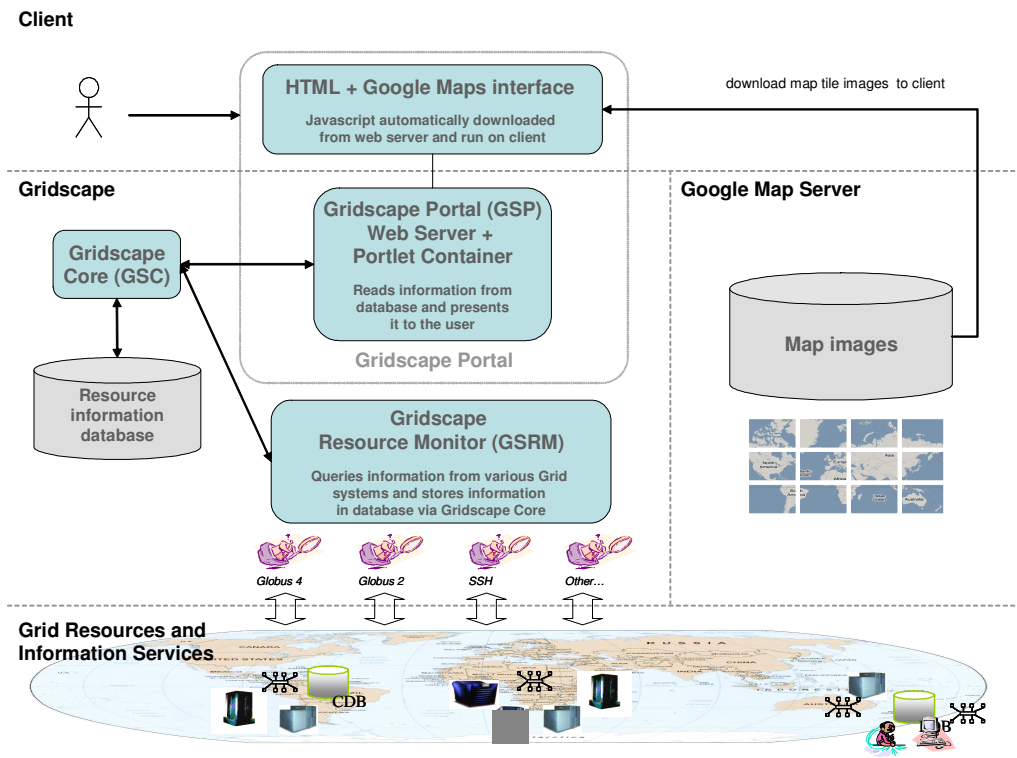

**Figure 1: Gridscape II Architecture**

The architecture of Gridscape II and its interaction with various Grid resources are illustrated in Figure 1. The figure identifies the three components which form Gridscape II: the Gridscape II Portal which provides the user interface, the Gridscape II Resource Monitor which gathers information about Grid resources, and the Gridscape II Core which is the data model for the system. The other important component is the Google Maps [16] JavaScript interface, which is a part of the portal.

## 3.1 GRIDSCAPE II CORE (GSC)

The Gridscape II Core (GSC) is the data model and represents the set of core components which are common to the rest of the system. This includes entities representing resources, location and resource information. The core also provides data access abstraction for accessing and managing Grid resource information. The overall structure of Gridscape II was aimed towards being integrated with a persistent backend so that the updating of resource information and the presentation of that information could be performed by entirely independent entities or separate applications, and so that state information is not kept in the memory of the presentation layer.

## 3.2 GRIDSCAPE II RESOURCE MONITOR (GSRM)

The Gridscape II Resource Monitor (GSRM) has the flexibility to be able to query any information source, and in a Grid context this includes Grid middleware such as Globus 2 and Globus 4 [17], Alchemi [18], Unicore [19], XGrid [20], SGE [21] and PBS [22]. Therefore, the GSRM is required to support a number of different protocols in order to gather information from these diverse information sources. The role of the GSRM is to gather information from these various sources and to publish that information for use by other components. The collective set of information becomes the current state of the Grid as seen by Gridscape II, and becomes available as a simple set of information to be utilised by a lightweight presentation layer. The importance of having this independent entity gathering resource information is that it alleviates the burden from the user interface and makes it easy to extend or create new user interfaces that are only required to present information. It was decided to avoid creating an information reporting agent that is installed on each resource, as this makes initial setup and adding new resources more difficult and often site administrators may not want to install such agents. As a result, Gridscape II can be used in a wide variety of applications and integrate with various information sources. Custom information gatherers can be created and plugged into the GSRM to enable the gathering of information from any type of information provider.

## 3.3 GRIDSCAPE II PORTAL (GSP)

The Gridscape II Portal (GSP) provides the user interface for the Gridscape II system. Using portlet technology, Grid resource information is presented to the user via their web browser. The portlet implementation also means that the interface can be easily plugged into other portals. To improve the interactivity of the otherwise static HTML interface, the Google Maps API is used to provide a highly interactive view of the globally distributed Grid resources and their positioning around the world. It is also important that the information is made available in a Web environment for increased accessibility and availability, as well as providing an intuitive user interface to enable easy navigation.

## 3.4 INTERACTIVE CLIENT-SIDE MAP (GOOGLE MAPS)

Google Maps provides an interactive world map on top of which other information, such as the location and properties of Grid resources around the world, is overlayed. The lightweight Javascript-based Google Maps client is downloaded onto the user's machine and runs within the web browser. From the web browser, asynchronous HTTP requests are made to retrieve map tile images from the map server that need to be presented as part of the portal. The asynchronous requests allow the user to zoom and pan around the map, while updating the display seamlessly without requiring an entire page refresh.

# 4 DESIGN AND IMPLEMENTATION

The overall design of Gridscape II is aimed towards being structured around enabling a persistent representation of resource information that can be updated without having to run a web server. Another key consideration is to allow the architecture to be flexible enough to support multiple different information sources and allow for easy integration of new source types and also provide a simple way to customise the presentation of the information without requiring changes to source code.

JSR168 [23] compliant portlets are used to implement the Gridscape II portal as a portable and pluggable component which can be integrated into other portals. The Gridscape II web application follows the Model-View-Controller (MVC) Model-2 type architecture [24]. This architecture provides a means of decoupling the logic and data-structures of the application (server-side business logic code) from the presentation components (web pages).

## 4.1 GRIDSCAPE II CORE

The GSC encapsulates the system state (see Figure 2) and acts as the model that represents the data and rules governing access and updates to this data. It consists of a Resource object representing a Grid resource and contains information collected about it in the ResourceInfo object. The Location represents the geographic location of that resource. The GSC is designed as a reusable component which can be used to develop other applications or views that need to work with Gridscape II data. The core is used by both the portal and the resource monitor.

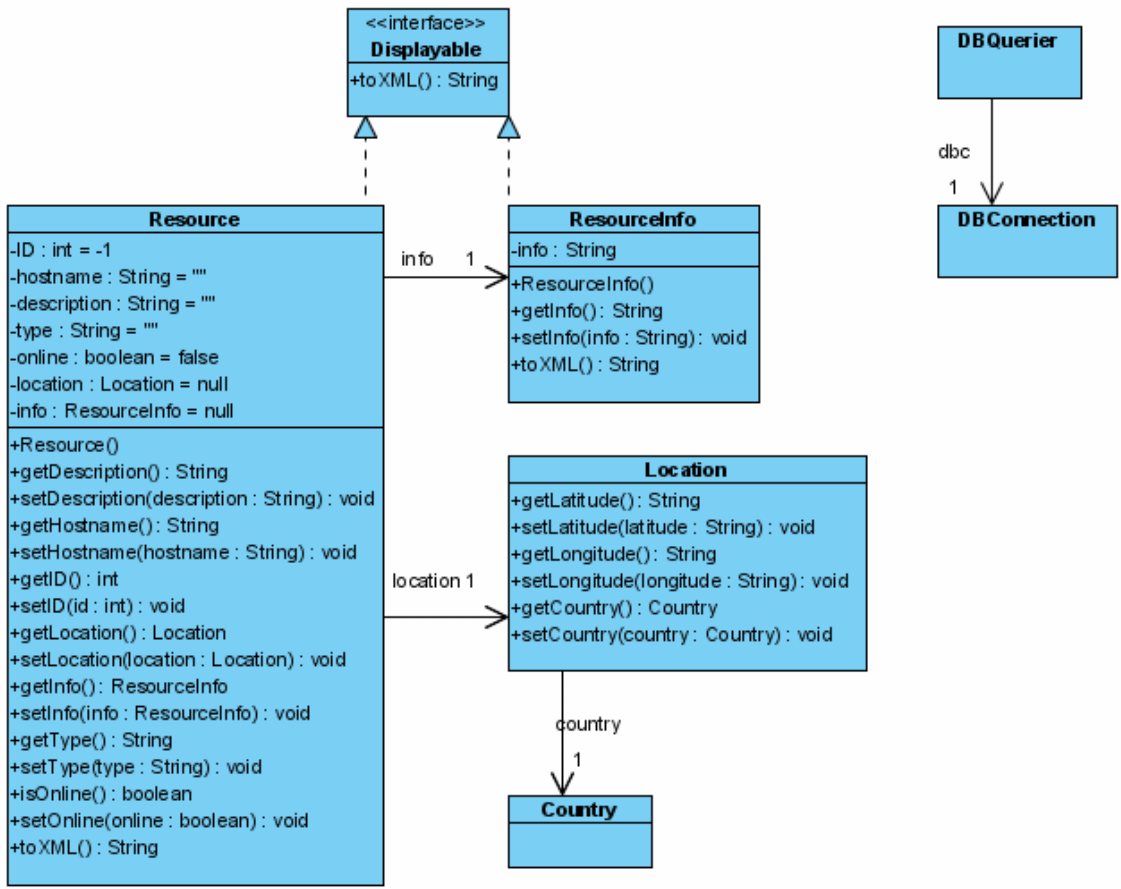

**Figure 2: Class Diagram of Gridscape II Core**

## 4.2 GRIDSCAPE II RESOURCE MONITOR

The GSRM, shown in Figure 3, is implemented as a standalone Java application. It consists of the ResourceMonitor class, which is the entry point to the GSRM, and the InfoGatherer interface along with implementations for each supported information service type. The GSRM is run independently of the GSP in order to gather, and publish via the GSC, state information from the various information service resources. When run, the ResourceMonitor will attempt to query each resource on a new thread and then sleep for a specified period of time before attempting to query resources again.

### 4.2.1    REFLECTION AND INFORMATION GATHERERS

An Information Gatherer is a plug-in which implements the functionality for gathering information from a certain type of information service. The GSRM queries the GSC for the list of resources and then for each resource, it uses the type property and reflection in order to load the Information Gatherer class that knows how to query that service. This design allows new types of information services to be introduced and used with Gridscape II by simply implementing a new plug-in (Information Gatherer) for that service.

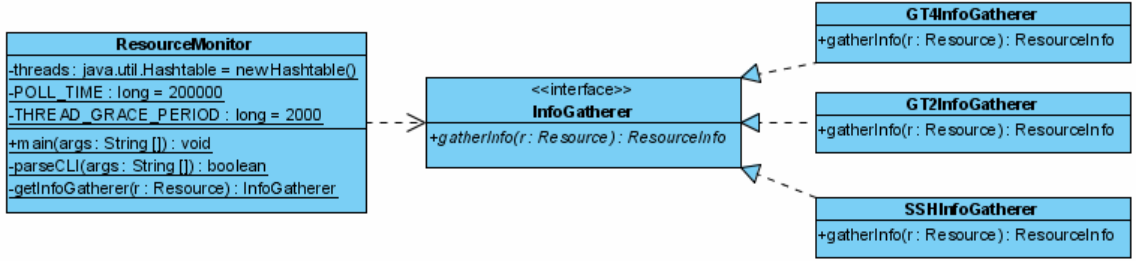

**Figure 3: Gridscape II Resource Manager Class Diagram**

### 4.2.2    IMPLEMENTING AN INFORMATION GATHERER

Introducing a new Information Gatherer is simple and only requires implementing the InformationGatherer interface which consists of a single method: gatherInfo(). The gatherInfo() method is required to query the resource and gather whatever information is necessary and return that information as a ResourceInfo object. The information returned is expected to be returned in XML format within the ResourceInfo object. This data will be stored to the GSC by the GSRM for use later during presentation.

## 4.3    GRIDSCAPE II PORTAL

Figure 4 shows the two individual portlets that make up the GSP - GridscapeEdit for administrating the Grid resources and portal appearance and GridscapeDisplay for presenting the resource information to users. The portlets control the program flow and prepare the objects to be presented by the JSPs. Each JSP utilises the Google Maps API and contains Javascript code to embed the interactive map interface. They also both use the StyleUtil class to format objects for display at runtime.

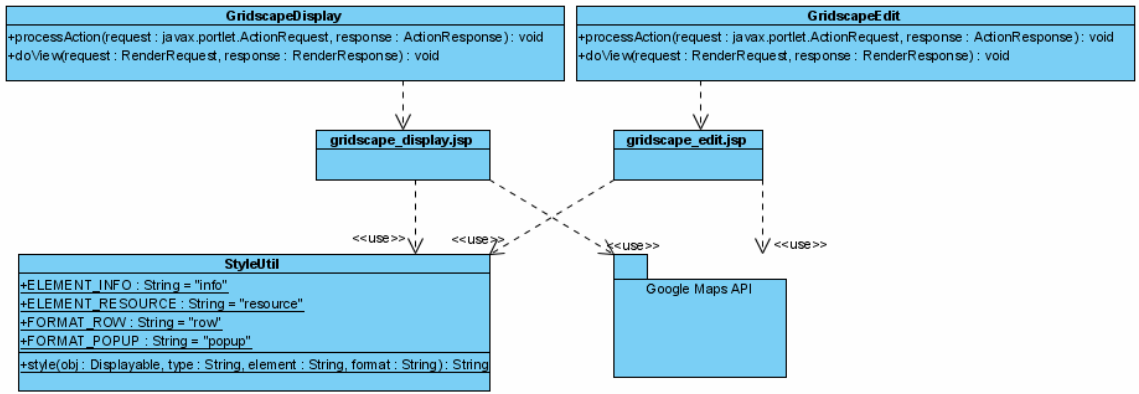

**Figure 4: Gridscape II Portal Class Diagram**

### 4.3.1    PRESENTING RESOURCE INFORMATION USING XSLT

eXtensible Stylesheet Language Transformations (XSLT) [25] is a way of transforming XML documents into other XML documents via stylesheets. In the case of Gridscape II, we are transforming the XML data gathered by the GSRM into HTML at runtime for presentation in the Web portal based on the appropriate stylesheet. The stylesheet describes how to access nodes of the source XML document using search patterns in the XPath language [26], and also how that should be transformed based on template tags to form the resulting document.

The benefit of this approach is the flexibility it provides. It means that we can allow the data for each information service to be stored based on any XML schema that is most suitable, and then simply requiring that an XSL document is available to transform that data into HTML that can be presented to the user. Currently Gridscape II requires that two stylesheets are available for each resource type, one for formatting the data to be displayed in the information window popup in the interactive map and another for displaying the information in the list view.

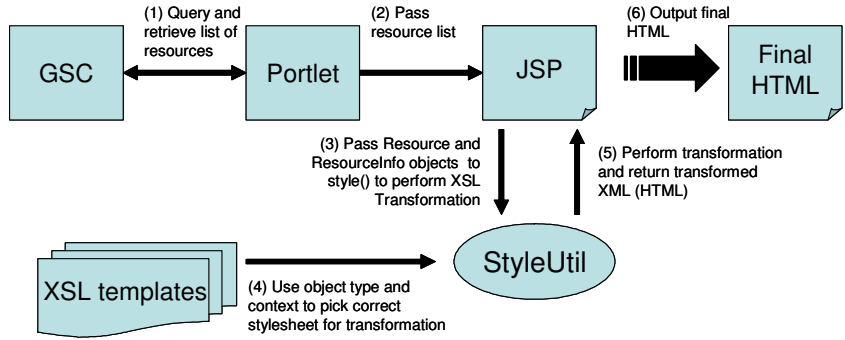

**Figure 5: Using XSLT to format objects for presentation**

The process of presenting resource information is shown in Figure 5. When a request comes in to render the portlet, the list of resources is queried and passed to the presentation layer – a JSP [27]. The JSP iterates through the list of resources to display each resource and makes a call to the helper class StyleUtil in order to perform the XSL Transform and format the information to be presented.

### 4.3.2    GOOGLE MAPS API AND AJAX

The Google Maps API is a free service offered by Google which allows developers to embed interactive maps (Google Maps) in their web sites. These maps display detailed mapping information and satellite imagery and allows interactivity such as panning and zooming of the map. The implementation of Google Maps is based on a web development technique that has recently become popular and has been termed Asynchronous JavaScript and XML or simply AJAX [28]. AJAX is used for making interactive web applications and aims to make web sites feel more responsive by sending asynchronous requests to a server in the background to retrieve small amounts of data rather than refreshing the entire web page. The way this works for Google Maps, as shown in Figure 6, is that when the map is displayed to the user, a request is made behind the scenes to a map tile server to retrieve the necessary tiles to be displayed, while the user drags the mouse on the map to pan the view, requests are made to the map server to fetch new tiles for the new region of the map to be displayed.

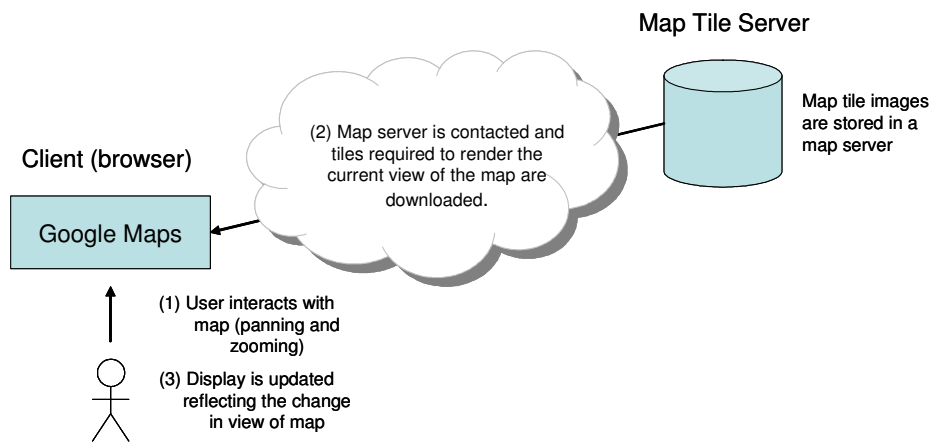

**Figure 6: Interacting with the Google Maps user interface**

The Google Maps API also allows developers to run overlays over their maps. These overlays include markers and paths. Markers are used to show points of interest in the form of an icon on the map. These icons are positioned based on latitude and longitude coordinates. They allow for the ability to pop up an information window above them upon user interaction. This capability is used in Gridscape II to indicate the position of Grid resources, and detailed information about each resource is made available in each marker's information window. Gridscape II dynamically generates JavaScript at runtime to produce markers for each resource. Paths are used to draw lines connecting points on the map. Paths are not yet utilised in Gridscape II.

## 5    GRIDSCAPE II IN PRACTICE

In this section we look at deploying, administrating and browsing the Gridscape II Portal.

### 5.1    CUSTOMISING THE PORTAL

Figure 7 shows the Gridscape II edit portlet which allows administration of Grid resources and general presentation details for your portal. Here resources can be added and removed and existing resources' details can be modified. The advanced options allow for customization of the interactive map including its size, the part of the world to focus on and whether users should be allowed to zoom and pan.

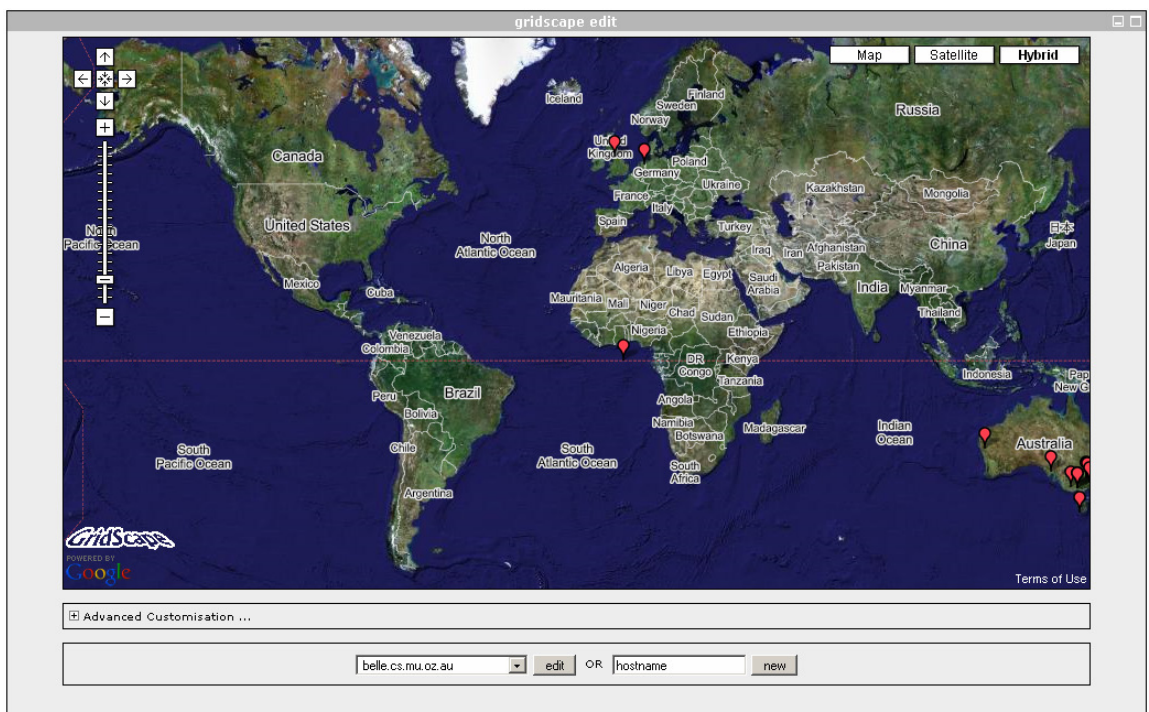

**Figure 7: Gridscape II Edit Portlet for portal administration**

#### 5.1.1    ADDING A NEW RESOURCE

Adding a new resource requires simply entering the hostname and then submitting the form. This will add the new resource to the list of existing resources and select it for editing so that its details can be entered.

#### 5.1.2    EDITING RESOURCE DETAILS

To change the properties of a resource, select the desired resource. This will present the properties of that resource in a form allowing them to be modified, as shown in Figure 8. The resource will also be shown as a green marker on the map. The hostname and type must be set correctly for the resources information to be successfully queried by the resource monitor. The location of the resource can be specified by clicking directly onto the map in the desired position. The map will update by displaying the new position.

**Figure 8: Editing resource details**

### 5.1.3 DELETING A RESOURCE

Resources that are no longer part of your Grid can be deleted from the system while selected for editing. This will permanently remove the resource, its marker and all other information that was gathered.

### 5.1.4 ADVANCED CUSTOMISATION

The advanced customisation options shown in Figure 9, allows the administrator to enter the Google Maps API key for the domain under which Gridscape II will run, which is a necessity when applying Google Maps to any Web site. The appearance and functionality of how the portal will be displayed to users can also be controlled here.

**Figure 9: Advanced Customisation options**

## 5.2 BROWSING THE PORTAL

Once customisation is complete the portal can be made available to users via the display portlet shown in Figure 10. The Gridscape II Resource Monitor should also be allowed to run so that the resource information can be gathered and updated for display in the portal.

### 5.2.1 INTERACTING WITH THE MAP INTERFACE

The map area shows the positions of all resources that have been entered into the Gridscape II portal. The user can interact with this by panning and zooming using the mouse and graphical interface controls. When a resource marker is selected, the information for that resource will be shown in an information window that will pop up above the marker. This information will be specific to the type of information source the user has selected.

### 5.2.2 SEARCH TOOLBAR

A search toolbar is available allowing users to search for specific resources using keywords. The search will look for any information or property that matches the keyword. Only those resources that are found to match will be displayed in the portal.

### 5.2.3    RESOURCE LIST

The resource list is displayed under the map and lists the properties and a summary of the resource specific information for each resource. This gives the user an overall view of the state of all the resources at a glance.

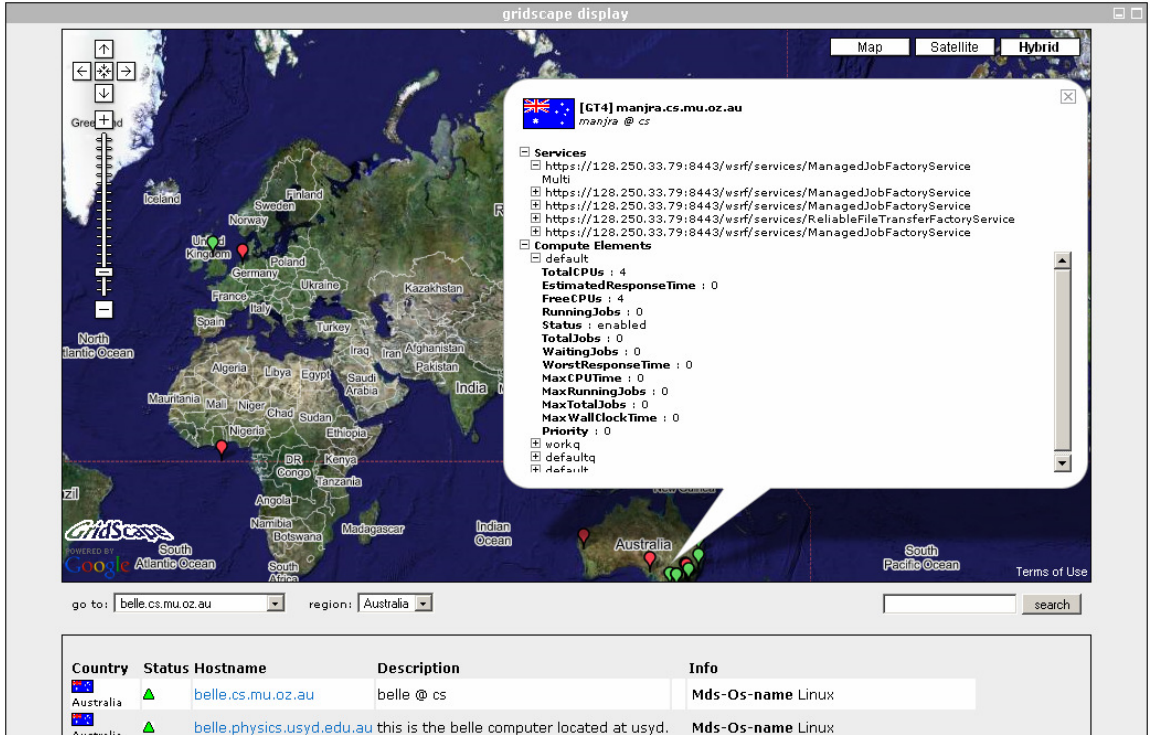

**Figure 10: Gridscape II Display portlet for viewing resource information**

## 6    CONCLUSION AND FUTURE WORK

With the overwhelming amount of information about Grid resources available from various different sources, tools such as Gridscape II, which provide a relevant and simple mechanism to navigate and view resource information, are required in order to improve our efficiency in working within such environments. There are currently a number of tools designed to monitor very specific details about a single type of resource or project, however, it is difficult for them to be customised and applied in other scenarios or Grids. Gridscape II thus provides a highly flexible solution that allows integration with current and future information sources, as well as providing an easy mechanism for customising what information is to be displayed and how it should be presented. Essentially Gridscape II can be applied in many situations and need not be restricted to any specific middleware, or Grid project or even to Grids in general.

We intend to add functionality to Gridscape II that allows it to monitor information regarding execution of Grid applications by integrating it with the Gridbus Broker [29]. We would also like to provide support for a number of existing resource types, such as Alchemi [18], Unicore [19], XGrid [20] or even providing access arbitrary nodes via SSH. Support for other types of resources such as Virtual Organisation systems [2] and data nodes may also be provided.

It is possible for resources to transmit their global positioning coordinates to Gridscape II while they are queried, and this is something currently defined in the GLUE schema [30]. Using this, the global positioning of resources could be more accurately plotted without the need for manual intervention.

Gridscape II is available at: http://www.gridbus.org/gridscape

# 7   ACKNOWLEDGEMENTS

We would like to acknowledge Google for inspiring the latest version of Gridscape with the release of its Google Maps API, and we would also like to thank Srikumar Venugopal for his thoughts on the implementation of Gridscape II.